\documentstyle[graphicx]{article}
\begin{document}

\title{Pioneer Anomaly and the Helicity-Rotation Coupling}
\author{John D. Anderson$^a$, Bahram
Mashhoon$^b$\vspace{.15in}
\\
$^a$Jet Propulsion Laboratory, California Institute\\ of Technology, Pasadena, CA 91109, USA\vspace{.15in}\\
$^b$Department of Physics and Astronomy, University\\ of Missouri-Columbia, Columbia, MO 65211, USA}

\maketitle

\begin{abstract}
The modification of the Doppler effect due to the coupling of the helicity of the radiation with the
rotation of the source/receiver is considered in the case of the Pioneer 10/11 spacecraft.  We explain why
the Pioneer anomaly is not influenced by the helicity-rotation coupling. 
\end{abstract}



\newpage

The Pioneer 10/11 Missions were launched about three decades ago and have been the first to explore the
outer solar system.  The analysis of radio tracking data since about two decades ago --- when the effective
acceleration away from the Sun due to the solar radiation pressure on the spacecraft decreased below
$5\times 10^{-8}\;{\rm cm/sec}^2$ --- has persistently indicated the existence of an anomaly in the
Doppler navigational data \cite{[1],[2],[3]}.  This Pioneer anomaly is a Doppler drift that may be interpreted as
being due to a small constant acceleration $a_P = (8.74\pm1.33)\times 10^{-8}\;{\rm cm/sec}^2$ of the
spacecraft {\it toward} the Sun.  The source of the anomaly is suspected to be a systematic effect
associated with the spacecraft; however, no definitive answer is known as yet \cite{[1],[2],[3]}.  

In a recent paper \cite{[4]}, the modification of the Doppler effect due to the coupling of photon spin with the
rotation of the emitter/receiver has been elucidated.  This result is of interest for the Doppler tracking
of spacecraft, since circularly polarized radiation is routinely employed to communicate with spacecraft
and, moreover, the Earth as well as the spacecraft rotate.  The question then naturally arises whether the
helicity-rotation coupling contributes to the Pioneer anomaly.  It is the purpose of this Letter to explain
why there is no such contribution, since the main effect has already been phenomenologically incorporated
in the analysis of Doppler data.  

Let $\omega$ and ${\bf k}$ be the frequency and wave vector of a photon according to inertial observers at
rest in a global inertial frame.  The photon is received by a noninertial observer with velocity ${\bf v}$
and rotation frequency $\mbox{\boldmath$\Omega$}$ such that $\Omega<<\omega$.  According to this rotating
observer, the frequency of the photon is 
\begin{equation}
\omega^{\prime} = \gamma\left[(\omega - {\bf\hat H}\cdot\mbox{\boldmath$\Omega$})-{\bf v}\cdot{\bf
k}\right]\;\;,\label{(1)}
\end{equation}
where $\omega = ck, \gamma = (1-v^2/c^2)^{-1/2}$ is the Lorentz factor and ${\bf\hat H} = \pm
c{\bf k}/\omega$ is the unit helicity vector of the photon.  The upper (lower) sign refers to a positive
(negative) helicity photon \cite{[4],[5]}.  The photon helicity is related to its intrinsic spin ${\bf S}$ by ${\bf S} =
\hbar{\bf\hat H}$.  A beam of positive (negative) helicity electromagnetic radiation is such that
inertial observers at rest along the beam looking down at the approaching wave will see the electric and
magnetic fields both rotate counterclockwise (clockwise) with frequency $\omega$ about the direction of
propagation.  Equation (\ref{(1)}) expresses the standard relativistic Doppler effect together with an extra term
due to the helicity-rotation coupling $(-\gamma{\bf\hat H}\cdot\mbox{\boldmath$\Omega$})$ that may be
neglected for $\Omega/\omega\rightarrow 0$; otherwise, for a small but nonzero $\Omega/\omega<<1$, ignoring
helicity-rotation coupling would lead to a systematic Doppler shift of magnitude $\Delta v =
c\Omega/\omega$ along the beam. 

It is possible to illustrate the contribution of helicity-rotation coupling to the Doppler effect by a
simple thought experiment.  Imagine an observer rotating uniformly with frequency $\Omega$ about the
direction of propagation of a plane monochromatic electromagnetic wave of frequency $\omega$ and definite
helicity.  Looking down at a positive helicity wave, the observer sees the electric and magnetic
fields rotate, relative to the observer, with frequency $\omega - \Omega$ about the propagation direction;
for a negative helicity wave, the relevant frequency would be $\omega + \Omega$.  Taking the time dilation
into account for the rotating observer, we find $\omega^{\prime} = \gamma(\omega\mp\Omega)$ in agreement
with equation (\ref{(1)}).  This result has been experimentally verified using the GPS system, where the
helicity-rotation coupling is known as the ``phase wrap-up'' \cite{[6],[7]}.  Further observational confirmations in
the microwave and optical domains are discussed in \cite{[4]}.  

In connection with the Pioneer Doppler drift, we note that the rotation frequency of the Earth
$\Omega_{\oplus}\approx 10^{-4}\;{\rm rad/sec}$ is much smaller than that of the Pioneer spacecraft, since
$\Omega\approx 0.4\;{\rm rad/sec}$ for Pioneer 10 and $\Omega\approx 0.8\;{\rm rad/sec}$ for Pioneer 11;
therefore, we shall first concentrate on the rotation of the spacecraft, which is essentially about the
direction of uplink radio beam from the Earth.  Let $\omega$ be the frequency of the uplink radio beam
with a definite helicity.  In the case of the Pioneer 10/11 spacecraft, $\omega/(2\pi)\approx 2\;{\rm GHz}$. 
Assuming that a spacecraft rotates with frequency $\Omega$ about the uplink direction, the frequency
received at the spacecraft is essentially $\omega_r = \omega\mp\Omega$ according to equation (\ref{(1)}), where the upper
(lower) sign refers to positive (negative) helicity radio waves. Here
we neglect terms of order $v^2/c^2\ll 1$ by setting $\gamma\approx 1$. Let us next assume that the spacecraft
transmits the same frequency $\omega_r$ back to the Earth --- though, in practice, the situation is more
complicated \cite{[2]} --- with the same helicity that was received.  Equation (\ref{(1)}) then implies that $\omega_r =
\omega_t\pm\Omega$.  Combining the two results, we find that the transmitted frequency received at the
Earth is $\omega_t = \omega\mp 2\Omega$.  
We therefore expect an anomalous spacecraft speed of
$2c\Omega/\omega$, which amounts to a few centimeters per second
for the Pioneer spacecraft, based on the supposition that the
helicity-rotation coupling has been totally neglected in the
analysis of Doppler data. That is, there would be a linear
phase drift as the phase wraps up, but the derivative of the
phase would be constant and hence there would be no anomalous
acceleration of the spacecraft. 

Let us next consider the rotation of the antenna that is fixed with
respect to the Earth. The rotation of the Earth would introduce an
anomalous spacecraft speed of $2c\Omega_\oplus \cos \theta/\omega$ as well, where
$\theta$ is the angle that the Earth-spacecraft direction makes with the
rotation axis of the Earth (colatitude). This anomalous speed is of order $10^{-4}$
cm/sec and is too small to be significant in the analysis of Pioneer
data \cite{[2]}. The angle $\theta$ in general varies with time; therefore, the
spacecraft would experience an anomalous acceleration of $-2c\Omega_\oplus
\dot{\theta} \sin \theta/\omega$, where $\dot{\theta} = d \theta/dt$. This
anomalous acceleration is negligibly small in the case of the Pioneer
spacecraft and hence does not affect the Pioneer anomaly. We now turn
to the explanation of how the main helicity-rotation coupling term has
in effect been taken into account in the analysis of Doppler data.   

Pioneer 10 was launched on 2 March 1972. In preparation for that
launch, the Navigation Team at JPL was concerned about two 
unpredictable effects---the characterization of the acceleration and
torques on the spacecraft from solar radiation pressure, and a
possible spin bias in the radio Doppler data used to determine the
spacecraft's Earth to Jupiter transfer orbit. The uncertainty in the
solar radiation effect was based on unknown absorptivities and
emissivities of spacecraft components and, to a lesser extent, their
effective surface areas. A solar-pressure model solved that problem by providing physical parameters
that could be inferred from the Doppler tracking data. For example,
shortly after launch the absorptivity and emissivity coefficients
determined from the tracking data provided a physically  reasonable
thermal model for both the front and back sides of the high-gain
parabolic antenna (2.74 m diameter). Those parameters, determined for
Pioneer 10 in 1972 and later Pioneer 11 after its launch on 5 April
1973, were used with confidence for the duration of the Pioneer
mission, up until termination of tracking in October 1990 for Pioneer
11 and March 2002 for Pioneer 10. 

The effect of the spacecraft spin on the Doppler data was not well
understood at launch. The high-gain antenna that would transmit the
radio carrier wave to Earth was attached to the spin-stabilized
spacecraft, and aligned with the spin axis, hence there was no doubt
that the antenna would be spinning at the same rate as the spacecraft,
and that the antenna would always be pointing approximately at
Earth. Based on a simple argument that the radio wave's electric
vector would be referenced to the rotating antenna, not inertial
space, the Navigation Team concluded that a constant frequency shift
would be introduced into the Doppler data, but experts in radio wave
propagation at JPL were not so sure. The propagation theory was
complicated by the fact that accurate tracking required a two-way
radio link with the spacecraft. There was no frequency reference on
board the spacecraft and it was necessary to do Doppler tracking by
means of a spacecraft transponder, with both uplink and downlink
transmissions referenced to atomic frequency standards at the
stations. It was known that the Deep Space Network (DSN) could transmit either left hand or
right hand circularly polarized radio waves, and it was also known
that at the spacecraft the transmission was right hand circularly
polarized from the cross-dipole antenna feed and left hand circularly
polarized from the parabolic dish. But there was no certainty in how
this information translated into a possible Doppler frequency bias,
and indeed more than one communications engineer predicted there would
be no spin effect in the two-way Doppler data. 

Faced with uncertainties about a possible spin bias, the Navigation
Team decided to apply a correction based on the idea that each
revolution of the spacecraft adds one cycle of phase to both the
uplink and downlink. The transponder ratio between the uplink
frequency at S-band, approximately 2.11 GHz, and the downlink
frequency was 240/221, hence it was decided to add a bias equal to $(1
+ 240/221)$ cycles per revolution of the spacecraft, or a frequency
bias in Hz of $(1 + 240/221)/P$, where $P$ is the spacecraft spin
period in seconds as measured by spacecraft sensors. However, even
given that the bias was appropriate, its sign was unpredictable. The
procedure shortly after launch was to apply the bias to the data, with
alternately a positive and negative sign and to see which sign gave a
better fit to the data. The positive sign definitely gave
a better fit, and further the alternative of no effect was ruled
out. Similar to the solar pressure constants, this inferred Doppler
bias was added to the frequency data for the duration of the
mission. Before 17 July 1990 it was added to the DSN data delivered to
the trajectory analyst. After that, it was not added to the DSN raw
data, and the analyst applied the correction. The Pioneer Project at
NASA/Ames Research Center provided the JPL analysts, by way of the
DSN, with the spin data for both Pioneer 10 and Pioneer 11. A bias
introduced by the rotation of the DSN antenna in inertial space was
never applied to the data. It is not clear if this was an oversight,
or if a calculation was made in the early 1970s and it was determined
that the DSN bias was buried at a negligible level in the Doppler
noise. 

\section*{Acknowledgment}
The work of J.D.A.\ was performed at the Jet Propulsion Laboratory,
California Institute of Technology, under contract with NASA.

\noindent Note Added: The determination of the Pioneer Doppler bias was made
even easier by the fact that the Pioneer 10 spacecraft was spun down
from 60 rpm to 5 rpm shortly after launch. The evolution of the bias
during this spin maneuver was obvious.

\begin{center}
\includegraphics[width=12cm]{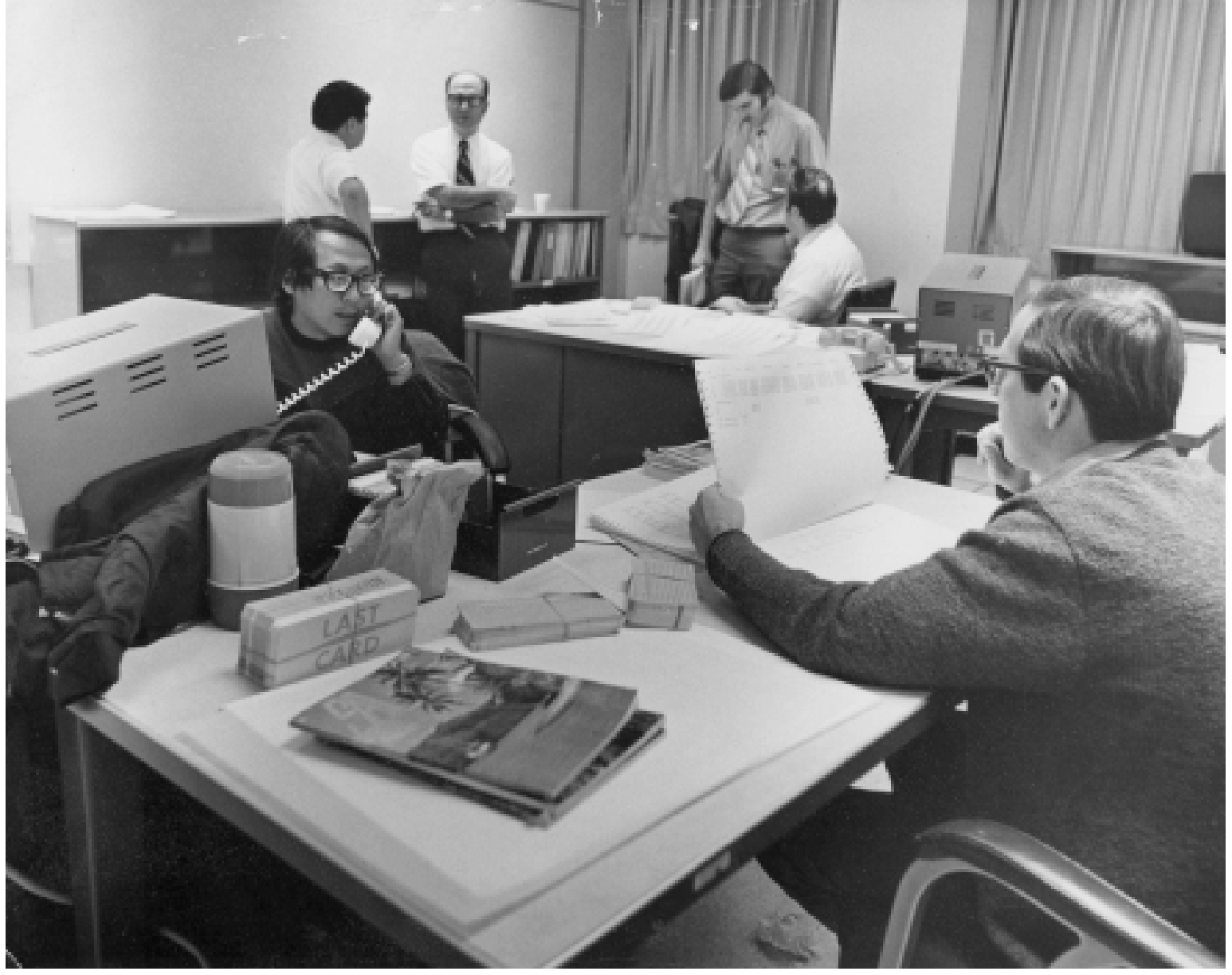}

\vspace{0.5cm}
\parbox{12cm}{The Pioneer Navigation Team in March 1972, during the launch phase of
Pioneer 10. Philip A.\ Laing, in the foreground, is poring over computer
output (Radio Science program POEAS) that will establish the sign of the
Doppler spin bias and verify its magnitude. A co-discoverer of the
effect, Anthony Liu, is on the telephone. The third co-discoverer,
Daniel L.\ Cain, is not present.}
\end{center}


\begin{thebibliography}{101}
\bibitem{[1]} J.D. Anderson et al., Phys. Rev. Lett. 81 (1998) 2858.
\bibitem{[2]} J.D. Anderson et al., Phys. Rev. D 65 (2002) 082004.
\bibitem{[3]} J.D. Anderson, M.M. Nieto and S.G. Turyshev, Int. J. Mod. Phys. D 11 (2002) 1545.
\bibitem{[4]} B. Mashhoon, Phys. Lett. A 306 (2002) 66.
\bibitem{[5]} B. Mashhoon, Phys. Lett. A 139 (1989) 103.
\bibitem{[6]} N. Ashby, in:  N. Dadhich and J. Narlikar (eds.), Gravitation and Relativity:  At the Turn of
the Millennium.  Proceedings of GR-15 (Inter-University Center for Astronomy and Astrophysics, Pune,
India, 1997), pp. 231-258.  
\bibitem{[7]} N. Ashby, Living Reviews in Relativity 6 (2003) 1.
\end{thebibliography}
\end{document}